\documentstyle[twoside,fleqn,espcrc2]{article}%
\newcommand{\beq}{\begin{equation}}
\newcommand{\eeq}{\end{equation}}
\newcommand{\bea}{\begin{eqnarray}}
\newcommand{\eea}{\end{eqnarray}}
\begin{document}                                                        
\renewcommand{\refname}{\normalsize\bf References}
\title{%
Dynamics of Complex Quantum Systems: Dissipation and Kinetic
Equations }

\author{%
        Aurel BULGAC%
        \address{Department of Physics,
                 University of Washington\\
                 P.O. Box 351560, Seattle, Washington 98195--1560, USA}%
\thanks{Address for the period October 1, 1999 -- June 30, 2000,
        Max--Planck--Institut f\"ur Kernphysik, Postfach 10 39 80,
        69029 Heidelberg, GERMANY.
        AB thanks H.A. Weidenm\"uller for being such a gracious host.
        The present work has been partially supported by DOE. 
        The Laboratoire de Physique Th\'eorique is a Unit\'e Mixte
        de Recherche du C.N.R.S., UMR 8627},
        \,Giu DO DANG%
        \address{Laboratoire de Physique Th\'eorique,
                 Universit\'e de Paris--Sud\\
                 B\^atiment 211, 91405, Orsay, FRANCE},%
        \,and
        Dimitri KUSNEZOV%
        \address{Center for Theoretical Physics,
                 Sloane Physics Laboratory\\
                 Yale University, New Haven, Connecticut 06520, USA}%
}
%
%
\begin{abstract}
\hrule
\mbox{}\\[-0.2cm]

\noindent{\bf Abstract}\\

We present a microscopic approach to quantum dissipation and sketch
the derivation of the
kinetic equation describing the evolution of a simple
quantum system in interaction with a complex quantum system. 
 A typical 
quantum complex system is modeled by means of parametric banded
random matrices coupled to the subsystem of interest.
We do not 
assume the weak coupling limit and allow for an independent dynamics
of the ``reservoir''. We discuss the reasons for having a new
theoretical approach and the new elements introduced by us.
The present approach incorporates known limits and previous results, but
at the same time includes new cases, previously never derived on a
microscopic level. We briefly discuss the kinetic equation and its
solution for a particle in the absence of an external field.   
\\[0.2cm]

\hrule
\end{abstract}

\maketitle

\section{Introduction}

The notion of a complex (quantum) system might not be an uniquely and 
universally accepted term, but it is certainly a very widely
(over)used one.  We shall call systems with a large number of
degrees of freedom complex systems. One might reasonable ask: ``Why a
new term?''. The real reason we believe, besides the temptation to
add something ``new'' to the jargon, is to make a distinction
with much simpler systems studied in the near past 
by the community contributing to this volume.
Typically, in complex quantum systems a
relatively small number of degrees of freedom (DoF) can be singled out in
one way or another and one might be interested, or forced, to study
the dynamics of this reduced set of DoF. Since these selected DoF
are coupled, either weakly, but most likely strongly, to the rest of
the DoF of the system and even to the rest of the Universe, one is
faced with the problem of devising ways to predict and
describe the dynamics of non--isolated quantum systems. 

In a certain
sense such systems can be also called open quantum systems. However,
the term open quantum system has a slightly different connotation. In
the case of an open system there is an ``orifice'' through which once
in a while the system can ``leak out'', or better put, ''drop out'', 
and never come back. In an open 
system the role of the ``rest of the Universe'' is typically reduced
to a simple ``black hole'' or a ``drain'' or a ``sink'', which  
``swallows'' matter out of the system as is ``been fed'', thus ``stuff 
simply falls out of the system'', 
and the ``rest of the Universe'' does not really exert any 
real force on the system of interest. One can say that the ``rest of the
Universe'' is ``forced fed'', but with no ``real effort''. 
The ``rest of the Universe'' can also serve as 
a ``source'' of matter, but again, this is done in the same somewhat
``non--obtrusive way'' as the time--reverse process of ``swallowing''.
The ``rest of the Universe'' can even play both roles, 
``sink'' and ``source'', at the same time.  
Thus an open system is nothing else, but a
system with somehow modified boundary conditions. One might argue that a
matter flux can influence dynamics. That would be correct if one would 
consider a system of interacting particles, whose number can vary in
time. The folks typically involved with open quantum systems study
physics at the independent particle level however.

A complex system in
the sense we shall be using is however somewhat different. We shall
not allow (yet) matter exchange, simply for the sake of simplicity,
but allow only for energy exchange between the
part of the system we have decided (or have been forced to) 
to focus our attention on and the ``rest of the Universe''.
Both parts of the system have their own dynamics, but at the
same time, the interaction between them could be strong enough as to
modify the dynamic evolution in a qualitative or quantitative
manner. The qualifier ``complex system'' refers as a rule either to
the ``rest of the Universe'' only or to the entire system. The dynamic
evolution of the ``rest of the Universe'' is routinely assumed to fall 
in one or another universality class, described in Quantum
Mechanics by some random Hamiltonian \cite{mehta}
or by some other Hamiltonian with ``complex dynamics''. 
If the interaction between the two subsystems would have been
weak, one would have dealt with a situation typically studied in
Statistical Physics textbooks. The energy exchange between the two subsystems
could be characterized as in Thermodynamics as either mechanical work
(thus reversible) or heat exchange (consequently irreversible). If
irreversible processes occur, one obviously has entropy production as well.
Moreover, if one would have concentrated one's 
attention only on the dynamics of the smaller system, the bigger one
would most likely have been called a thermal bath, and one would have
aimed at obtaining a kinetic description of the dynamics of the
smaller/simpler system. If the ``rest of the Universe'' is huge, one
might reasonable argue that the influence of the subsystem on it
should be negligible and thus could be ignored. Besides the fact that
that is not always the case of interest (when the ``rest of the
Universe'' is not huge, like in the case of the interaction between
the collective DoF in a nucleus and the rest), there is no need
to influence all the ``rest of the Universe'', but
only the part lying nearby and if the heat conductivity of the ``rest
of the Universe'' is low, energy will not be quickly
transferred to the outskirts of the ``rest of the Universe''. 

There was a significant interest over the years in quantum problems
similar to the type we have just described. Caldeira and Leggett \cite{cl},
using a technology outlined by Feynman and Vernon \cite{fv} have
studied the motion a ``heavy'' particle interacting with a reservoir
of harmonic oscillators. The reservoir of harmonic oscillators is
assumed to mock a quantum ``heat'' reservoir. The quantum Hamiltonian 
describing the ``Universe'' according to Caldeira and Leggett has the form
\bea
H&=&\frac{P^2}{2M}+U(Q) \nonumber \\
&+&\sum_{k=1}^\infty \left [
  \frac{p_k^2}{2m_k} +\frac{m_k\omega _k^2(q_k-c_kQ)^2}{2}\right ], 
\label{eq:cl} 
\eea
with the corresponding $p$'s and $q$'s and so forth and the coupling
constants $c_k$. A continuous distribution of these harmonic oscillators 
leads to a dissipative dynamics of the ``heavy'' system described by
momentum $P$ and coordinate $Q$ and in particular to a Langevin
description in the classical limit. The fact that this 
particular classical limit should emerge in the high--temperature
regime, was one of the requirements imposed on the structure of the
model Hamiltonian in Eq. (\ref{eq:cl}). This model Hamiltonian suffers
however from one rather simple inadequacy, which is at least
intellectually unsatisfactory. Let us assume that we would like
to describe the dissipative motion of a particle in a homogeneous
medium. There is  no external potential $U(Q)$ then and it
is completely unclear in this case how one can define a coupling to
the reservoir, as there is obviously no special point with respect to
which one can measure $Q$, since the medium is supposed to be
homogeneous. One might consider instead a coupling through a derivative 
term, i.e. to the momentum of the ``heavy'' particle. Then one would
have had to alter the time--reversal properties of the Hamiltonian
for no other particular reason. A coupling to the momentum typically
implies some gauge field. In particular, such a mechanism will not work 
in one dimension, as no magnetic field exists in 1D. Moreover, simply
having to consider a gauge field for the sake of having dissipation
looks like an overkill, at least from a theoretical point of
view. Coupling to spin degrees of freedom for the same reason should
be ruled out as generic. Thus we seem to find ourselves in a quandary.
Even in the presence of an external field, the fictitious 
harmonic oscillators become more and more elongated as the ``heavy''
particle moves away from the ``center'' and thus the ``reservoir''
becomes ``infinitely stretched'' and consequently, due to the presence 
of an infinite number of DoF in the reservoir, ``infinitely
excited'' too. Moreover, even though it is entirely 
reasonable to be guided by the ``known'' classical limit when
``guessing'' the structure of the model Hamiltonian, one can hardly
accept this as a satisfactory recipe and a suitable solution for all
cases. On one hand, the form of the friction force is not unique even
in classical physics, classical friction force can be proportional to
the velocity at various powers or even independent of velocity. On
the other hand, atomic nuclei for example represent systems in which there is
no classical limit and they are certainly not unique in this
respect. Moreover, there very well could be a genuine quantum
dissipative regime, which has no counterpart in the classical limit. 

We would like to start from a model ``typical Hamiltonian'' of a
``typical quantum complex system'', with a minimum and relevant amount 
of input and find out under what conditions a dissipative dynamics emerges, 
what is its character and how various properties of this dynamics
depend on the particular properties of the initial Hamiltonian. It is
not obvious that one will arrive at the same type of description as
Caldeira and Leggett have obtained. As a matter of fact we find that
choosing this alternative approach one sometimes arrives at a
dissipative dynamics similar to that found by Caldeira and
Leggett. The surprise however is that dissipative dynamics with
entirely new characteristics emerges as well. The particular example of 
quantum L\'evy flights and fractional kinetics is one remarkable case 
\cite{kbd99}.

There is another direction of inquiry in the theory of open quantum
systems or quantum dissipative dynamics, which we shall mention only very 
superficially \cite{l}. The main emphasis in these approaches is
phenomenological, at the level of the von Neuman equation and its 
possible extensions, such as to accommodate a Markovian evolution, to
preserve the positivity of the density matrix and to conserve the probability. 
As one might have expected from a pure phenomenological approach, 
there has been no effort to understand the nature and the
origin of various terms added to the von Neuman equation from any
underlining microscopic dynamics, as no such derivation has ever been 
attempted, except cases when perturbation theory could be implemented and one 
ends up with a textbook master equation.

This sketchy review of the literature is not meant to be
exhaustive or comprehensive in any sense. The aim was merely to point
the reader to some literature leads and to justify in some very 
limited way our approach and line of reasoning. We deeply apologize
for not devoting more space and emphasis to so much fine work, 
not mentioned here mostly due to space limitations. 
The rest of the contributors to this volume will surely 
 complement us.

\section{A generic quantum complex system }

We shall describe here what we view as a generic quantum complex
system and the reasons why we have chosen this perspective. The
Hamiltonian of the ``entire Universe'' we choose as follows \cite{bdk98}
\beq
{\cal{H}}=\frac{P^2}{2M} + U(Q) + H(Q) ,
\eeq
where $H(Q)$ is supposed to describe the ``rest of the Universe'' and
the interaction with the selected DoF with canonical coordinates
$(P,Q)$, which could stand for one or more pairs of canonical variables. 
The ``rest of the Universe'' is assumed typically to be a
very large system, with a large heat capacity. It is convenient to
choose an average level density of states of the ``rest of the Universe''
which is locally exponentially increasing, 
\beq
\rho(\varepsilon )=\overline{\mathrm{Tr} \delta (H(Q)-\varepsilon )}=
\rho_0\exp (\beta \varepsilon ).
\eeq 
The overline stands for averaging over some portion of the spectrum or 
for an ensemble averaging to be described shortly.
Thus $\beta=1/T $ is nothing else but the inverse thermodynamic
temperature of the ``rest of the Universe''.
So far we have always chosen an average level density
which is independent of the coordinate $Q$
\cite{kbd99,bdk98,bdk97,bdk96}. In physical terms this means that only
heat exchange is allowed so far between the selected subsystem and the 
``rest of the Universe''. We do not see any difficulties of any kind
to incorporate the possibility for the system to perform work on the
``rest of the Universe'' as well (or vice versa). Allowing for the temperature 
$T$ to vary apparently could require some further small technical 
developments of our formalism. The relevant ``dependence'' of the
Hamiltonian $H(Q)$ on the subsystem coordinate $Q$ is implemented by
modeling it with a random matrix \cite{mehta} and 
specifying the second cumulants of its matrix elements:  
\cite{bnw}
\bea
\!\!\!\!\!\! & &\overline{[H(X)]_{kl}[H(Y)]}_{mn}
-\overline{[H(X)]}_{kl} \quad \overline{[H(Y)]}_{mn} \nonumber \\
\!\!\! &=&[\delta_{km}\delta_{ln}+\delta_{kn}\delta_{lm}] 
\frac{\Gamma ^\downarrow }{2\pi \sqrt{\rho (\varepsilon _k)
\rho (\varepsilon _l)}}  \nonumber \\
\!\!\! & \times & \exp \left [ 
-\frac{(\varepsilon _k -\varepsilon _\l)^2}{2\kappa _0
^2} \right ]G \left (\frac{X-Y}{X_0}\right ).
\eea
Here $\Gamma ^\downarrow$ is the so called spreading width, $\kappa
_0$ defines the bandedness of the random matrix and it can be
interpreted as the energy of a ``single kick''.
To $X_0$ we refer to
as the correlation length. Even though we have used GOE--type of
induced spectral fluctuations, similar results are obtained for the
GUE or GSE case. This in a way could have been expected. 
It is natural to expect that most/many complex systems have
universal spectral fluctuations at some small energy scales as both
theory and experiment indicate. 
A GOE Hamiltonian is
suitable for an even number of fermions, while a GSE Hamiltonian for an
odd number of fermions \cite{mehta}. 
A ``reservoir'' with an odd or even number of
fermions could not influence in qualitatively different ways a given 
subsystem.

The correlation length $X_0$ is perhaps one of the most
interesting parameters of the entire theory. The function
$G((X-Y)/X_0)$ ($G(0)=1$ and $G(x)=G(-x)=G^*(x)\le 1$)
is assumed to have a bell shape with a characteristic width $X_0$. Thus
$X_0$ specifies by how much one has to move the subsystem $(Q,P)$ in
order to induce some noticeable changes in the ``rest of the
Universe''. Obviously, if the ``rest of the Universe'' does not react
in any way and does not change its properties while the subsystem
$(Q,P)$ evolves, there is no energy exchange and no change in
entropy as well. 
(We trust that this is neither the place nor the time where one should 
discuss at any length why entropy can vary, even though one might claim
that the entropy of the entire system is strictly conserved.) The
restriction adopted here of ``translation invariance'' has no
fundamental consequences and apparently can be lifted without much effort.

From the characterization given so far of the Hamiltonian $H(Q)$ of
the ``rest of the Universe'' and of the coupling to the subsystem $(Q,P)$ 
one can see the differences with other approaches mentioned in the
introduction, in particular with the most ``microscopic'' theory of 
Caldeira and Leggett \cite{cl}. The ``translation invariance'' of the
present approach includes the case of the motion of a subsystem $(Q,P)$ in a
homogeneous medium.

From this point on the development of the theory follows Feynman
and Vernon  strategy \cite{fv}, one writes down a double path--integral
for the density matrix of the entire system, one integrates out the
``rest of the Universe'' and in the end one arrives at an evolution equation
for the density matrix of the subsystem $(Q,P)$. It is surprising that 
in the limit of high temperatures and large bandwidth of the matrix 
$[H(Q)]_{kl}$
(namely, when $\kappa _0\rho(\varepsilon ) \gg 1$) one arrives and an
extremely simple looking evolution equation for the reduced density
matrix of the subsystem $\rho (X,Y,t)$, which reads:
\bea
\!\!\! &\!\!\! &i\hbar \frac{\partial \rho (X,Y,t)}{\partial t}  =\left\{ 
  \frac{P_X^2}{2M}-\frac{P_Y^2}{2M}+ U(X)-U(Y)  \right. \nonumber  \\
\!\!\! &\!\!\! &
  -\frac{\beta \Gamma ^{\downarrow }\hbar }{4X_0M}
  G^{\prime }\left( \frac{X-Y }{X_0}\right) (P_X-P_Y)  \nonumber \\
\!\!\! &\!\!\! &+\left. 
  i\Gamma ^{\downarrow }\left[ G\left( \frac{X-Y}{X_0}\right)
  -1\right] \right\} \rho (X,Y,t).  \label{eq:evol}
\eea
Hopefully no confusion with the notation
used earlier for the average level density should arise.
The derivation, which is somewhat lengthy and requires an extensive
discussion of various steps, is described in our earlier works and
shall not be repeated here. The ``rest of
the Universe'' is allowed to have its own dynamic evolution and it
is not assumed to be in thermal equilibrium or to have a prescribed
evolution. Some rather innocent assumptions are also made about the
nature of the initial conditions \cite{bdk98,bdk96,bdk95}.
In the case when one simply drives the ``rest of the
Universe'' one can show that the energy distribution has
significant non--Maxwellian distortions \cite{bdk96}. The coupling
between the subsystem $(Q,P)$ and the ``rest of the Universe'' is not
assumed to be weak at any stage, even though the limit of weak
coupling is obviously embodied as well in the evolution equation 
(\ref{eq:evol}).

By means of a Wigner transform and in the case when for small
arguments 
\beq
G(x)\approx 1 -x^2/2 + \ldots  \label{eq:smallx}
\eeq
the evolution equation (\ref{eq:evol}) reduces in the limit 
$\hbar \rightarrow 0$ to the
Kramers equation \cite{bdk98}, and thus the equivalent Langevin
description is also obtained. The classical friction force in this
case is proportional to the first power of the velocity, the
friction and diffusion coefficients are given by
\bea
\gamma &=&\frac{\beta \Gamma ^\downarrow \hbar}{2MX_0^2},\label{eq:g} \\
D_{QQ} &=&\frac{2X_0^2}{\beta ^2 \Gamma ^\downarrow \hbar} \label{eq:D}
\eea
and the standard Einstein dissipation--fluctuation relation is
satisfied.

One might have considered the fact that this classical limit is
exactly reproduced under apparently correct conditions as a necessary 
self--consistency check of the entire approach. The power of the present 
approach is demonstrated however by being able to generate
other distinct and consistent classical limits as well, which, as far as 
we can judge, was not shown to be possible within the framework of 
approaches previously known in the literature. In particular, a rather 
innocently looking change of the correlator $G(x)$ for small arguments to 
\beq
G(x)=1-|x|^\alpha +\ldots 
\eeq
with $0<\alpha \le 2$ leads to apparently the first microscopic
derivation of fractional kinetics with L\'evy flights \cite{kbd99},
without assuming or introducing any kind of exotic noise. The only
``noise'' is of the usual rather boring Gaussian type.
A completely
different behavior, turbulent--like diffusion \cite{kbd97}
\beq
\langle \langle Q^2\rangle \rangle \propto t^3
\eeq
is also in the realm of this type of evolution equation
(\ref{eq:evol}). The double angle brackets stand for cumulants.

Surprisingly, in many cases the evolution equation can be solved in 
quadratures for arbitrary type of coupling between the two subsystems 
\cite{kbd99,bdk98,bdk97}. Time--dependent solutions of the 
Schr\"odinger equation are known for the case of free motion, linear
potential, harmonic potential. Path--integrals
also can be calculated in cases when the action is essentially
quadratic in $p$'s and $q$'s. 
To our surprise the solution of Eq.(\ref{eq:evol}) can be
obtained easily if the classical trajectories in the absence of
dissipation (i.e. no $G((X-Y)/X_0)$) are known. Several cases were
discussed by us in Refs. \cite{bdk98,bdk97}. Here we shall
illustrate a few salient points of our approach on the example of
motion in the absence of any external field ($U(Q)=0$), but for
arbitrary type of coupling to the ``rest of the Universe''. In 1D 
and representing the density matrix in the coordinates $r=(X+Y)/2$,
$s=X-Y$ one can show that for an arbitrary initial density matrix
$\rho _0(X,Y)$, for $t>0$ one has 
\bea
& & \rho (r,s,t) = \nonumber \\
& & \int\!\int \frac{dr^{\prime }dk}{2\pi \hbar }\rho
_0\left ( r^{\prime },s-\frac{kt}{M}\right ) \exp \left[ 
\frac{ik(r-r^{\prime })}\hbar \right.
\nonumber \\
& &\left. +\frac{\Gamma ^{\downarrow }M}{\hbar k}\int_{s-kt/M}^sds^{\prime
}[G(s^{\prime }/X_0)-1]\right] .  \label{rho}
\eea
This form of the density matrix is especially suited to evaluate
the coordinate and momentum cumulants as functions of time.

In the long time limit the momentum distributions reaches an
equilibrium, characterized by vanishing odd cumulants, while the even
cumulants are given by ($n\ge 2$)
\bea
\langle\langle  P^{2n}\rangle \rangle =
(-1)^{n-1}\frac{(2n-1)!!}{n} \frac{MX_0^2}{\hbar ^2 \beta } \left (
    \frac{\hbar}{X_0}\right ) ^{2n} , \!\!\!
\eea
for the particular case of a Gaussian correlator
\beq
G(x)=\exp(-x^2/2).
\eeq 
The second cumulant has the expected value
\beq 
 \langle\langle  P^2   \rangle \rangle = 2MT ,
\eeq
if only for small arguments the correlator behaves as 
in Eq. (\ref{eq:smallx}).
There is nothing special about the Gaussian form of the correlator $G(x)$ used 
here, which
was simply chosen for illustrative purposes.
The parameter which governs the deviation of the equilibrium
distribution from the one expected in the weak coupling limit Maxwellian
distribution is $\hbar/X_0$ and not as one would have naively guessed 
the ``strength'' of the coupling $\Gamma ^\downarrow $.
$\hbar/X_0$ could be interpreted as the characteristic momentum exchanged
between the system and the ``rest of the Universe'' in ``one completed 
interaction act''. The equilibrium momentum distribution is narrower
than a pure Maxwellian distribution at the same temperature.

Some information about the coordinate distribution has been inferred
as well \cite{bdk98}. The second cumulant reveals what one would have
expected, namely that for $t\rightarrow \infty$
\beq
\langle\langle  Q^2   \rangle \rangle \propto 2D_{QQ}t ,
\eeq
again, if only for small arguments the correlator behaves as 
in Eq. (\ref{eq:smallx}).
A qualitative analysis of the coordinate distribution seem to indicate 
the presence of ``longer tails'' than one would have obtained in the
case of a simple Brownian particle. That does not seem so surprising
now when we have learned that under very mild circumstances one
easily obtains  L\'evy flights behavior and fractional kinetics 
\cite{kbd99} or turbulent--like diffusion \cite{kbd97}. 
However, a comprehensive analysis of the entire
coordinate distribution is not achieved yet. 

Analytical results can also be obtained for a particle 
in a linear potential or in a quadratic potential (both harmonic
oscillator and parabolic barrier) and qualitatively similar deviations 
from the textbook cases of the weak coupling regime are found. The very 
interesting case of tunneling in a double well potential has been
studied only numerically so far \cite{bdk98,bdk97}, along with other
cases which have some resemblance to under the Coulomb barrier collision
and tunneling of two atomic nuclei. Evidently, our
approach is not limited to 1D as one might have concluded, simply
because we did not explicitly display spatial subscripts. 

\section{Concluding remarks}

On a closer analysis and in retrospective the fact that we have used a 
random matrix ensemble, with various adds-on, does not seem to have
played a fundamental role. If that would have been the case we would
have seen some distinction between GOE, GUE and GSE cases, which we
did not confirm \cite{bdk95}, 
even though at one time that was expected \cite{w}, but later on
apparently not confirmed as well \cite{wa}. The fact that a 
random matrix approach is not critical to obtaining the final answer
was confirmed as expected in Ref. \cite{lw} in the weak coupling
limit. As we have mentioned in the Introduction, it would  have been
indeed very strange if a GOE--reservoir would lead to a different
kinetic equation than a GUE-- or GSE--reservoir. 

However, the particular approach based on the use of parametric banded 
random matrices allowed us significantly more freedom in choosing
various forms of coupling between the ``rest of the Universe'' and the 
subsystem of interest. In the wide band limit of a random matrix we
find that the kinetic equation is determined essentially completely by 
the correlator $G(X-Y)$, which defines correlation between properties
of the Hamiltonian describing the ``rest of the Universe'' at two
different locations of the simple subsystem. Besides the thermodynamic 
temperature of the ``rest of the Universe'' one needs essentially only 
two more quantities to define the character of the dissipative
dynamics, the ``intensity'' of the coupling controlled by the
spreading width $\Gamma ^\downarrow$ (we avoid using the term strength
of the coupling for reasons discussed in Section  2) and by the so
called correlation length $X_0$, which defines the typical momentum exchanged
between the two subsystems. In the usual $\hbar \rightarrow 0$ limit,
the friction and diffusion coefficients depend explicitly on
$\hbar$, see Eqs. (\ref{eq:g},\ref{eq:D}), which seems
inconsistent. The resolution of this apparent puzzle is simple, 
one cannot simply
consider the limit $\hbar \rightarrow 0$, while keeping $X_0$ and
$\Gamma ^\downarrow$ fixed; the corresponding combinations entering
in these relations could be held fixed, in particular 
$\Gamma ^\downarrow \hbar /X_0^2$. 

As we have seen, the freedom of choice our approach gives us by
allowing various shapes for the correlator $G(X-Y)$, leads to new
interesting cases, such of those of L\'evy flights dynamics and
turbulent diffusion, and perhaps some other cases, which at the moment 
we do not suspect yet.

Even though the limit of a wide band ($\kappa _0\rho (\varepsilon )\gg 
1$) could prove more then adequate for many practical applications, it
would be satisfying to attempt the calculation of the influence
functional using alternative techniques, like the super--symmetry method 
\cite{vwz}, which do not rely on the $1/N$--expansion. We did mention
in previous publications that the present approach seems amenable to
describe localization as well, when the finite bandwidth of the random 
matrix is explicitly taken into account. In that case the influence
functional emerging from this model has the structure discussed in
Ref. \cite{c}. 

It would also be interesting to determine under what general requirements one
can derive such kinetic equations, in particular by not using
random matrix models at all. In spite of the fact that random matrix
models have proved by now to be ``universal'', we have seen that to
some extent even they seem to be too specific.

\end{document}